\documentclass[reprint,aps,prb,superscriptaddress,showpacs]{revtex4-1}
\usepackage{graphicx}
\usepackage{dcolumn}
\usepackage{bm}
\usepackage{amsmath}
\usepackage{amssymb}
\usepackage[]{units}
\usepackage{booktabs}
\begin{document}

\title[]{\textit{Ab-initio} friction forces on the nanoscale: A DFT study of fcc Cu(111).}

\author{M Wolloch}
\email{mwo@cms.tuwien.ac.at}
\affiliation{Institute of Applied Physics, Vienna University of Technology, Gu\ss hausstra\ss e 25-25a, 1040 Vienna, Austria}
\affiliation{Austrian Center of Competence for Tribology, Viktor-Kaplan-Stra\ss e 2, 2700 Wiener Neustadt, Austria}
\author{G Feldbauer}
\affiliation{Institute of Applied Physics, Vienna University of Technology, Gu\ss hausstra\ss e 25-25a, 1040 Vienna, Austria}
\affiliation{Austrian Center of Competence for Tribology, Viktor-Kaplan-Stra\ss e 2, 2700 Wiener Neustadt, Austria}
\author{P Mohn}
\affiliation{Institute of Applied Physics, Vienna University of Technology, Gu\ss hausstra\ss e 25-25a, 1040 Vienna, Austria}
\author{J Redinger}
\affiliation{Institute of Applied Physics, Vienna University of Technology, Gu\ss hausstra\ss e 25-25a, 1040 Vienna, Austria}
\author{A Vernes}
\affiliation{Institute of Applied Physics, Vienna University of Technology, Gu\ss hausstra\ss e 25-25a, 1040 Vienna, Austria}
\affiliation{Austrian Center of Competence for Tribology, Viktor-Kaplan-Stra\ss e 2, 2700 Wiener Neustadt, Austria}

\begin{abstract}
While there are a number of models that tackle the problem of calculating friction forces on the atomic level, providing a completely parameter-free approach remains a challenge. Here we present a quasi-static model to obtain an approximation to the nanofrictional response of dry, wearless systems based on quantum mechanical all-electron calculations. We propose a mechanism to allow dissipative sliding, which relies on atomic relaxations. We define two different ways of calculating the mean nanofriction force, both leading to an exponential friction-versus-load behavior for all sliding directions. Since our approach does not impose any limits on lengths and directions of the sliding paths, we investigate arbitrary sliding directions for an fcc Cu(111) interface and detect two periodic paths which form the upper and lower bound of nanofriction. For long aperiodic paths the friction force convergences to a value in between these limits. For low loads we retrieve the Derjaguin generalization of Amontons-Coulomb kinetic friction law which appears to be valid all the way down to the nanoscale. We observe a non-vanishing Derjaguin-offset even for atomically flat surfaces in dry contact.
\end{abstract}


\maketitle

\section{Introduction}
\label{sec:Int}

In the $17^{th}$ and $18^{th}$ century Amontons and Coulomb formulated the basic laws of friction, which state that the macroscopic friction force is independent of the apparent contact area and varies linearly with the external load at moderate constant sliding velocities \cite{amonton:1699,coulomb:1773}. These friction laws have been further extended by Derjaguin for adhesive surfaces, so that a non-vanishing friction force can occur even for zero load \cite{derjaguin:34b}. Although generally very successful, on the nano-scale deviations from this classical behavior have sometimes been observed depending on the sliding conditions. If the contacting surfaces are aligned incommensurately, friction can be reduced by several orders of magnitude and structural superlubricity may occur \cite{hirano:90,hirano:97,dienwiebel:04,dietzel:13,zhang:13}, though this phenomenon is suspected to be unstable at least for graphene flakes on graphite \cite{filippov:08, feng:13}. It has also been proposed that the drastic reduction of frictional forces may not stem from incommensurability but from thermal effects and the low effective mass of the nanocontact \cite{krylov:05,maier:05}. Inclusion of third bodies between incommensurate contacts hinders structual superlubricity and leads to a linear behavior of friction on load \cite{he:99,mueser:01}, which is also recovered from thermodynamic considerations \cite{gao:04}. Other studies report a greater than linear \cite{gosvami:10}, or sublinear dependence of the friction force on load \cite{wenning:01,luan:05,mo:09} where the latter is consistent with classic Hertzian contact mechanics \cite{hertz:1881}. Also even more complex dependencies of the friction force on the load have been reported~\cite{eder:11,vernes:12}. A more general overview on friction simulations on the nanoscale may be found in two recent reviews \cite{dong:13,vanossi:13b}.
Friction forces can also depend on the sliding direction as shown in experiments on rather complex geometries~\cite{overney:94,park:05}, and recently by Weymouth et al.~for a single atom asperity \cite{weymouth:13}.

While the conventional method of choice for simulating atomic scale friction is classical molecular dynamics (MD), see Ref.~\onlinecite{dong:13} and references therein, \textit{ab-initio} density-functional-theory (DFT) apporaches \cite{hohenberg:64,kohn:65} have become more common during recent years \cite{garvey:11a,garvey:11b,garvey:11c,zibilotti:11,cahangirov:12,lf_wang:12,j_wang:12,kwon:12}. This is due to advantages in accuracy and the increasing power of computers which retain the simulation time manageable. In classical MD simulations, energy becomes dissipated to a heat bath via a thermostat, a method which is in principle also possible for DFT calculations. As simulation times for \textit{ab-initio} MD are very short and thermostating is difficult due to a rapid heating of the electronic system, it is desirable to formulate a different way to describe dissipative sliding and to predict coefficients of friction within DFT. This may be done by computing potential energy landscapes and fitting the resulting energy barriers to mechanical models, for example the well known Prandtl-Tomlinson model \cite{prandtl:28,tomlinson:29}. Another approach was suggested by Zhong and Tom\'{a}nek, who, in their ``maximum-friction microscope'' model, assume a complete dissipation of the potential energy into phonons and electronic excitations for each slip~\cite{zhong:90,tomanek:91}.

Considering all these different concepts and methods, we attempt to formulate the problem of dry sliding friction on the basis of parameter-free (\textit{ab-initio}) calculations, which rely only on the quantum-mechanical interactions of the sliding bodies. To this end we propose energy dissipation via the relaxations in the sliding materials themselves as calculated \textit{ab-initio}. We will show that the concept allows to analyse arbitrary sliding directions and sliding paths up to $\mu m$ scale in length with a single initial set of DFT calculations. This permits us to gain insight into the different behavior of periodic and aperiodic (defined below) sliding directions by bridging four orders of magnitude in length scale.

\section{Methods}
\label{sec:Meth}

\begin{figure}[htbp]
\centering
\includegraphics[width=0.51\linewidth]{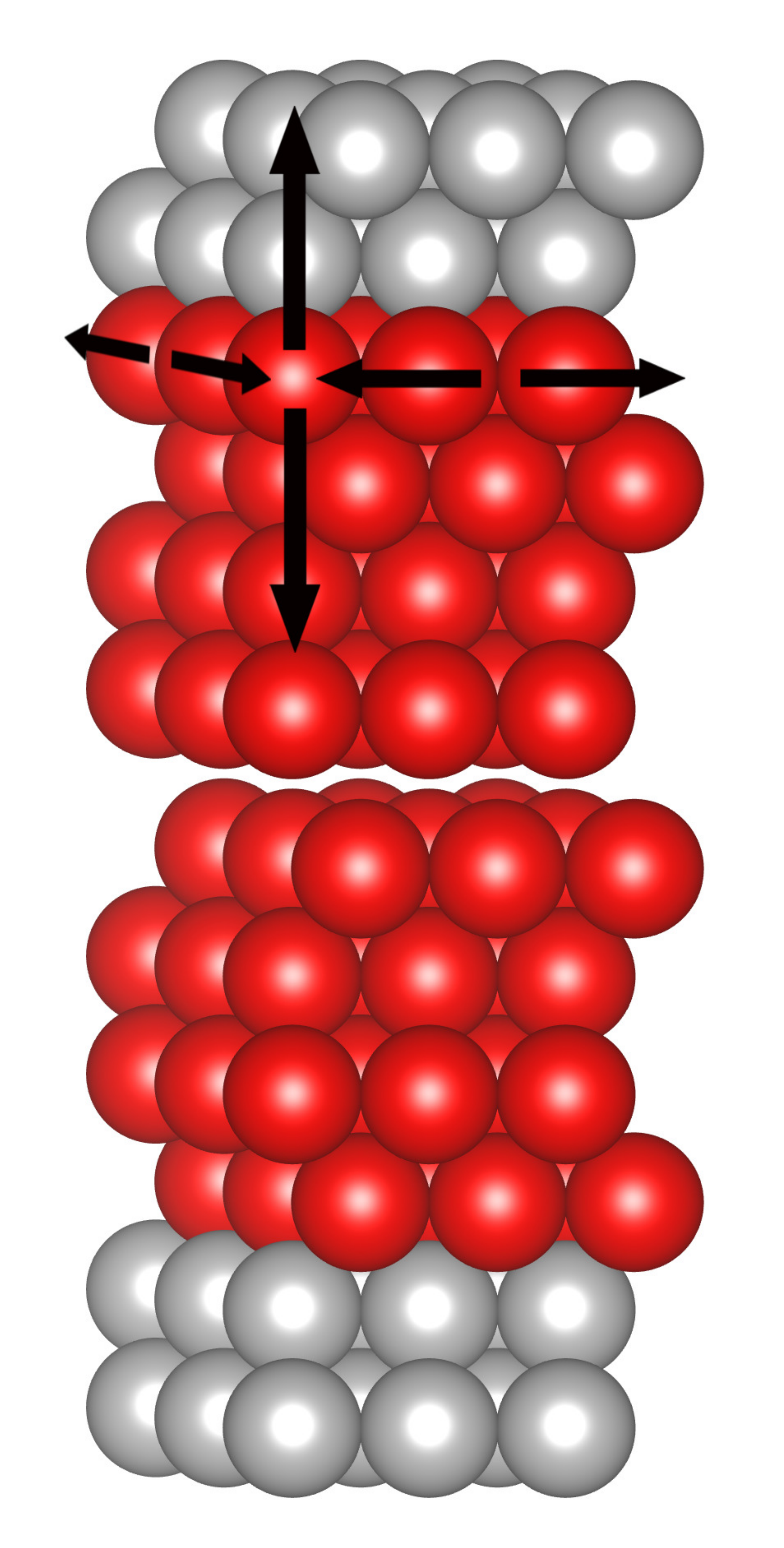}
\caption{Sketch of the simulation cell containing 12 Cu atoms, tripled in the x and y direction for clarity. The top slab is displaced vertically to simulate several load conditions and in the x and y direction to collect data on energies and forces. Red atoms are allowed to relax, while grey atoms are kept rigid at their bulk-like positions. The gap at the interface is included for visual clarity and is of course not present under load when the slabs are compressed beyond their equilibrium distance.}
\label{fig:0}
\end{figure}

As a prototype system to study our nanofriction model we consider two atomically flat slabs of fcc(111) copper in dry contact, represented by a $1\times 1$ hexagonal unit cell (see Fig.~\ref{fig:0}). Both slabs are described by two rigid Cu layers (grey spheres) representing the transition to Cu-bulk followed by a tribologically active zone consisting of four Cu layers (red spheres) which are allowed to relax in 3 dimensions. In preliminary calculations the addition of an additional free layer was found to not significantly change the resulting geometries even at the positions with the highest stresses. External loads are imposed by keeping the rigid layers at a given distance smaller than the equilibrium one. The full simulation box consists of 12 atoms using a bulk lattice constant $a=\unit[3.634]{\AA}$ determined from DFT equilibration, cf. in experiments $a=\unit[ 3.615]{\AA}$ \cite{wyckoff:63}. Due to the inherent periodicity of our supercell, a vacuum layer of about \unit[15]{\AA} is included on top of the upper slab to decouple the periodically repeated simulation cell in the z direction. This 12 layer thick arrangement is comparable to a number of previous studies using \textit{ab-initio} methods~\cite{dag:04, neitola:06, wieferink:11, zhang:11, garvey:11a, garvey:11b, garvey:11c, cahangirov:12, j_wang:12, cahangirov:13}, and is sufficiently thick to approximate the elastic properties of copper while retaining computational efficiency.

\begin{figure}[htbp]
\centering
\includegraphics[width=1.0\linewidth]{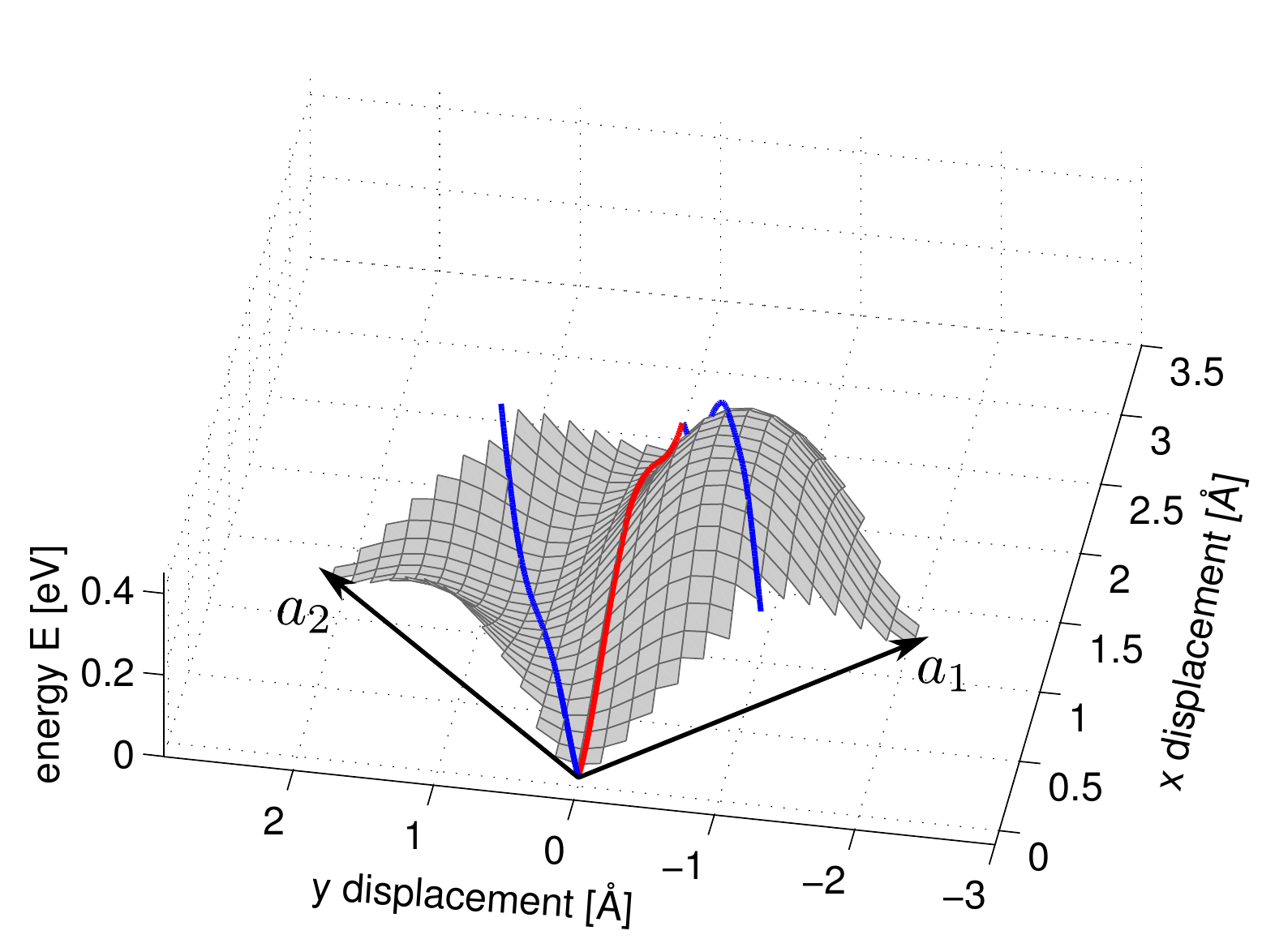}
\caption{(color online) Energy landscape of a constant load scan for two fcc Cu(111) slabs displaced
  relativ to each other. In addition, two periodic sliding paths with $0^\circ$
  (red) and $30^\circ$ (blue) are shown. The angles are defined with respect to the Cartesian x-direction. The arrows $a_1$ and $a_2$ are two basis vectors of the rhombohedral unit cell. The x-direction dissects the plane spanned by $a_1$ and $a_2$ and points into the $[\,1\,1\,\overline{1}\,]$ direction. The y-direction points into the $[\,\overline{1}\,1\,0\,]$ direction and is in the plane spanned by $a_1$ and $a_2$ and orthogonal to the x-axis.}
\label{fig:1}
\end{figure}

To investigate nanofrictional forces for sliding paths of different lengths and directions we apply a quasi-static grid approach, which means that the unit cell is sampled by choosing points on a regular grid for which the energies and forces are calculated (see Fig.~\ref{fig:1}). In addition to the advantage of being a parameter-free method, this concept ensures that the sliding direction can be chosen arbitrarily without recalculating energies and forces. The same holds for the length of the sliding path which can easily be extended to $\sim\unit[1]{\mu m}$, four orders of magnitude larger than the dimensions of the unit cell. To calculate the energies and forces for each lateral position a $10\times10$ grid is constructed in the x-y plane, resulting in a spacing of $\unit[\sim0.25]{\AA}$ between the grid points. A cubic spline interpolation is applied to refine this grid by a factor of 10 and create smooth energy- and force-surfaces.
To simulate the usual experimental setup, we perform our scans keeping the load at a constant value which implies that the distance of the slabs needs to adapt. To this end the distance between the slabs is varied in 6 steps for each of the 100 grid-points, recording the respective volume $V$ and the total energy of the relaxed cell $E^{\mathrm{R}}$ which are fitted to a second order polynomial. The uniaxial pressure on the cell can now be evaluated by calculating the first derivative of $E^{\mathrm{R}}$ with respect to $V$ at each lateral position, namely $p=-\partial E^{\mathrm{R}}\,/\,\partial V$. The corresponding loading force $L$ is then obtained by multiplying the pressure with the cross-section area $A$ of the unit cell, $L=p\,A$. Choosing a load (given derivative of the energy vs. volume functions) we recalculate the respective energies and forces at each of the 100 grid points and obtain energy surfaces at that given constant load. Evidently that means, that the distance between the slabs has to be adjusted accordingly. Employing this process, which was also used in a very similar way by Cahangirov et al.~\cite{cahangirov:12}, we construct quantum mechanical energy- and force-maps for the quasi-static sliding system, both with and without relaxation of atoms.

It has to be pointed out that this model is not intended to simulate a macroscopic copper on copper system which would feature multiple grains, oxidation, impurities, and other imperfections. The aim is rather to study a model for dissipative sliding under idealized and controlled conditions. We believe that the modeled system and the computational method are ideally suited for this endeavor for the following reasons: i) commensurate sliding of a Cu(111) tip on Cu(111) was previously found to exhibit wearless atomic stick-slip motion, both in molecular dynamics simulations \cite{sorensen:96,martini:09}, and experiments in ultra-high vacuum~\cite{bennewitz:99}. This is in contrast to results for Cu(100) on Cu(100) where both plastic deformations and wear play an important role \cite{sorensen:96,nieminen:92}. ii) Garvey et al. showed in a recent series of papers on KCl sliding on Fe(100) that the investigation of shearing interfaces and the prediction of friction coefficients for such systems require very high accuracy in the calculated energies \cite{garvey:11a,garvey:11b,garvey:11c}. The Vienna Ab-Initio Simulation Package \textit{VASP}~\cite{kresse1993,kresse1994a,kresse1996a,kresse1996b} applied to the copper system seems to be ideally suited for our study. The Projector Augmented Wave (PAW) code~\cite{bloechl1994,kresse:98} is much more accurate as for example embedded atom potentials but not as computationally expensive as highly accurate full-potential linearized augmented plane-wave (FLAPW) method which is needed sometimes for systems where the local bonding environment changes significantly \cite{garvey:11a}.
We found that, while the considered system is small enough that the large number of calculations needed for the construction of accurate energy surfaces for various loads is still feasible, it is also realistic enough to work as a proof of concept for the proposed scheme to calculate friction forces on the nanoscale.

One of the main goals of this work is to examine the differences in nanofriction for different sliding directions, especially the differences between periodic and quasi aperiodic sliding paths. For all sliding directions we start in an energy minimum configuration corresponding to perfect fcc stacking. Different paths are distinguished by their angle with respect to the x-axis, see Fig.~\ref{fig:1}. When a sliding path exits at a unit cell boundary, the underlying lattice periodicity demands that it reenters on the opposite side. In Fig.~\ref{fig:1} we show two periodic paths with 0 and 30 degrees, respectively. The first (red path) in Fig.~\ref{fig:1} dissects the unit cell once along the Cartesian x-axis before returning to its starting point, whereas the second one (blue path) in Fig.~\ref{fig:1} cuts the unit cell twice. As can be seen from Fig.~\ref{fig:1} the energy variations along these two paths are distinctly different representing two extremal cases. Additional to these periodic paths we also consider 10 quasi aperiodic ones, namely for $-33$, $-29$, $-15$, $-10$, $3$, $5$, $10$, $21.13$, $45$, and $50$ degrees. We note that due to the inevitable representation of irrational numbers by rational fractions in a computer these paths are essentially also periodic, however, with very long periods. The forces acting on the upper slab have a parallel (friction) and an in-plane perpendicular (reaction) component with respect to the direction of the path. Since our model allows for both positive and negative contributions to the friction force, we use the convention that friction forces pointing against the sliding direction for a given sliding path are counted as positive.

As already mentioned all energies and forces are calculated with the DFT package \textit{VASP} using the PAW method with an energy cut-off of \unit[341.5]{eV}. For the exchange and correlation potential, the Generalized Gradient Approximation (GGA) in the version devised by Perdew, Burke and Ernzerhof (PBE) is applied~\cite{perdew:97}. As the strong chemical bonds in copper are well described with PBE and our results rely on the calculation of total energies, there is no need to explicitly consider van der Waals forces or meta-GGA. The Brillouin zone sampling is performed on a $\Gamma$-centered $24\times24\times1$ k-grid ensuring a total energy convergence better than \unit[1]{meV} per simulation cell. Atomic relaxations were converged to \unit[0.1]{meV} in total energy, with each electronic calculation being converged to \unit[0.01]{meV}.

\begin{figure}[htbp]
\centering
\includegraphics[width=1.0\linewidth]{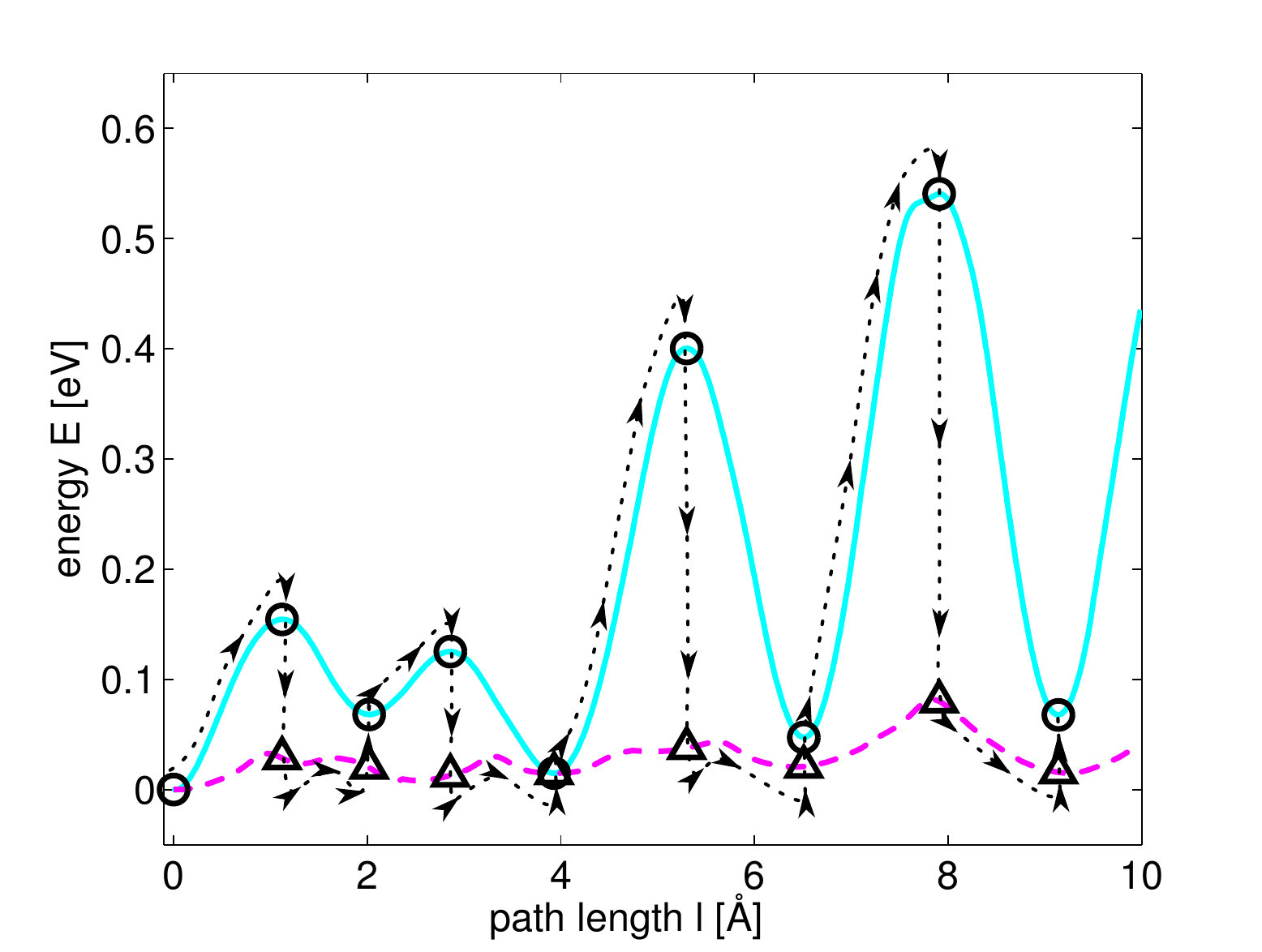}
\caption{Illustration of the energy dissipation model along a $50^\circ$ sliding path. The solid (cyan) line gives the energy for the unrelaxed slabs with the open (black) circles denoting local minima and maxima. The dashed (magenta) line shows the energy for the relaxed slabs with open (black) triangles marking the positions of the extrema of the unrelaxed slabs. The dotted (black) line is a schematic route for the calculation of the energy differences between the open circles and triangles, see equation~(\ref{eq:F2}).}
\label{fig:2}
\end{figure}

A central feature in the description of friction is to allow dissipative sliding. In principle, any modeling that involves smooth continuous energy surfaces, such as we get from our grid approach when using only the fully relaxed system, means that one moves on a conservative energy landscape, resulting in energy neutral displacements and frictionless sliding. An earlier \textit{ab-initio} attempt to tackle this problem has been developed by Zhong and Tom\'{a}nek et al.~\cite{zhong:90,tomanek:91} who studied the stick-slip motion of single atom tips over graphite. In contrast to their approach and in order to describe the portion of energy lost into heat, we propose the following mechanism. Along each path we identify the local minima and maxima on the energy curve of the unrelaxed system. We assume static sliding without relaxations until a maximum in the energy is encountered, which in our model is representative of the ``stick'' phase of stick-slip sliding. At this point we allow the built up strain to be released by relaxing the tribologically active zone (the first part of the ``slip'' phase). The resulting energy difference is now assumed to be dissipated into the bulk crystal via phononic excitations and is ultimately lost as heat. Although the method is quasi-static the resulting frictional forces can be predicted sufficiently well, if the amount of energy that is lost during sliding is estimated correctly. The relaxed energy surface is also corrugated, thus, the movement to the next minimum may result in small gains or losses in the energy leading to small negative or positive friction forces along this portion of the path. This is the second part of the ``slip'' phase. When the next minimum in the unrelaxed surface is encountered the cycle is repeated until the desired path length is reached. If the local minimum of the unrelaxed energy curve is not at the same energy as the relaxed curve, an appropriate portion of the released energy from the last slip is used to bring the system back to the unrelaxed sliding position. With this method we avoid the use of springs to model the elasticity of the system, which is instead included implicitly in the relaxations. The process is visualized in Fig.~\ref{fig:2}.

\begin{figure}[htbp]
\centering
\includegraphics[width=0.9\linewidth]{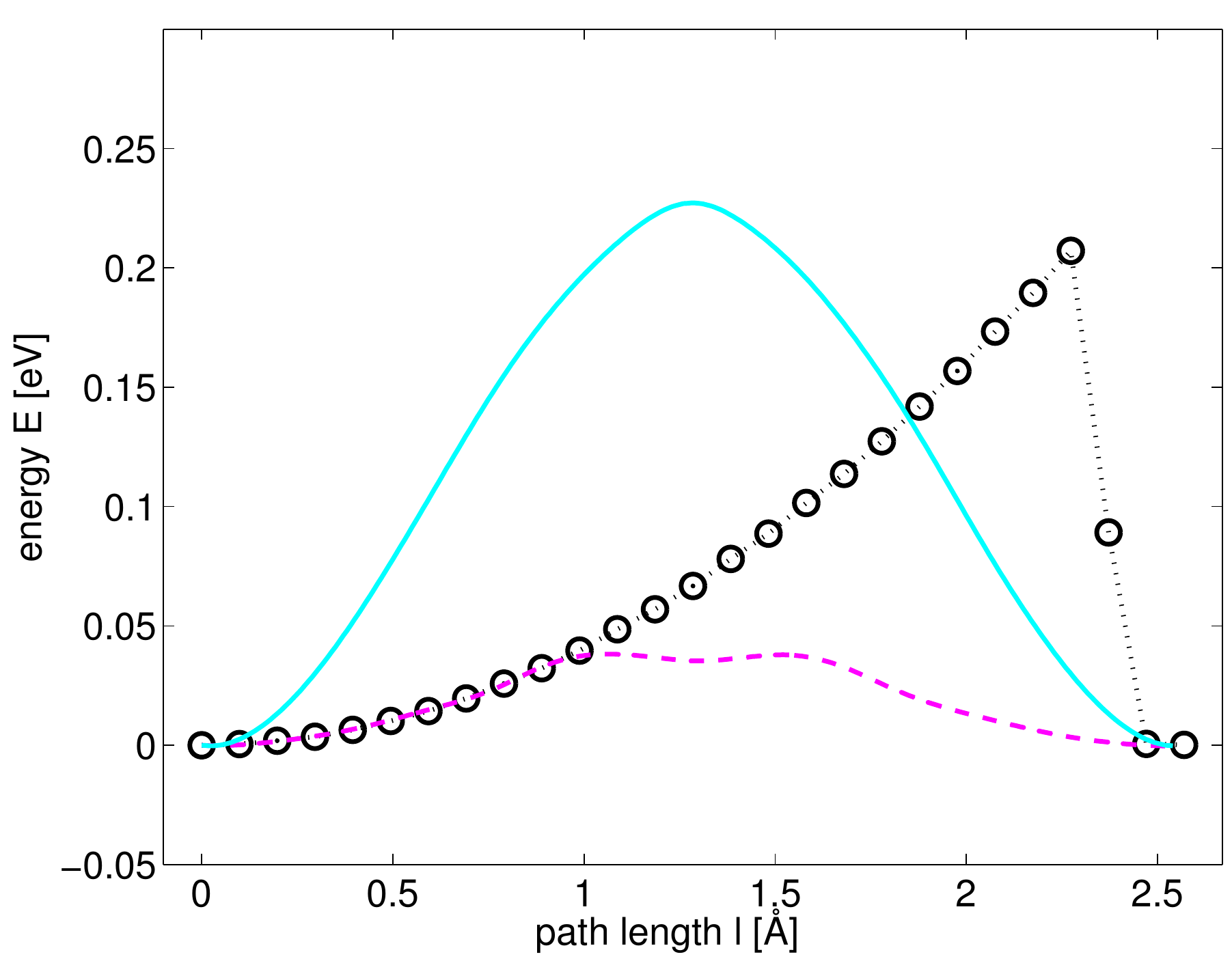}
\caption{A comparison between our friction model and a shearing simulation for the periodic $0^\circ$ sliding path at zero load. The entire period is shown. The solid (cyan) line and the dashed (magenta) line give the unrelaxed and relaxed energy curves for the quasi-static model, as seen also in figure~\ref{fig:2}. The black circles are fully relaxed data points of the shearing calculation.}
\label{fig:3}
\end{figure}

The use of unrelaxed energy curves, where the two bodies slide over each other statically, only adjusting their vertical distance to keep the load constant, is of course a rather crude assumption. However, for the two periodic sliding directions at $0^\circ$ and $30^\circ$ we also carried out more realistic shearing calculations that support our approach. In these calculations we use the same unit cell and drag the uppermost two (fixed) layers in small ($\sim\unit[0.1]{\AA}$) steps along the chosen sliding direction while holding the two bottom-most layers still. At each step the 8 free layers in the middle are allowed to relax and, in contrast to the quasi-static approach, these relaxed positions are the starting configuration for the next step. This method is only computationally practical for short path lengths and becomes extremely time consuming for long aperiodic paths. The vertical distance between the two fixed regions of the slab was kept constant during these calculations and at the equilibrium distance of the non-sheared slab, i.e.~corresponding roughly to zero load. The results for the $0^\circ$ path are shown in figure~\ref{fig:3}. 

In the beginning the energy of the sheared slabs follows the relaxed energy curve from our proposed quasi-static model, but starts to deviate in the vicinity of its first maximum at about $l=\unit[1]{\AA}$. The sheared system is not fully relaxing into the total energy minimum but is still pinned in the potential well of the starting position, a common feature in stick-slip sliding. The slip is initiated only at the end of the period, where the built up stain energy of $E_{\mathrm{shear}}(0^\circ)=\unit[207]{meV}$ is dissipated and the system slips into the next minimum. This energy is actually underestimated by $\sim 25\%$ within our proposed quasi-static friction model at $E_{\mathrm{qs}}(0^\circ)=\unit[156]{meV}$ although we include contributions of the unrelaxed energy curve (see figure~\ref{fig:2} and equation~(\ref{eq:F2})). In contrast, applying the model by Zhong and Tom\'{a}nek~\cite{zhong:90,tomanek:91} to the relaxed energy curve underestimates the dissipated energy during shearing by more than 80\% at $E_{\mathrm{ZT}}(0^\circ)=\unit[38]{meV}$.

\begin{figure}[htbp]
\centering
\includegraphics[width=0.9\linewidth]{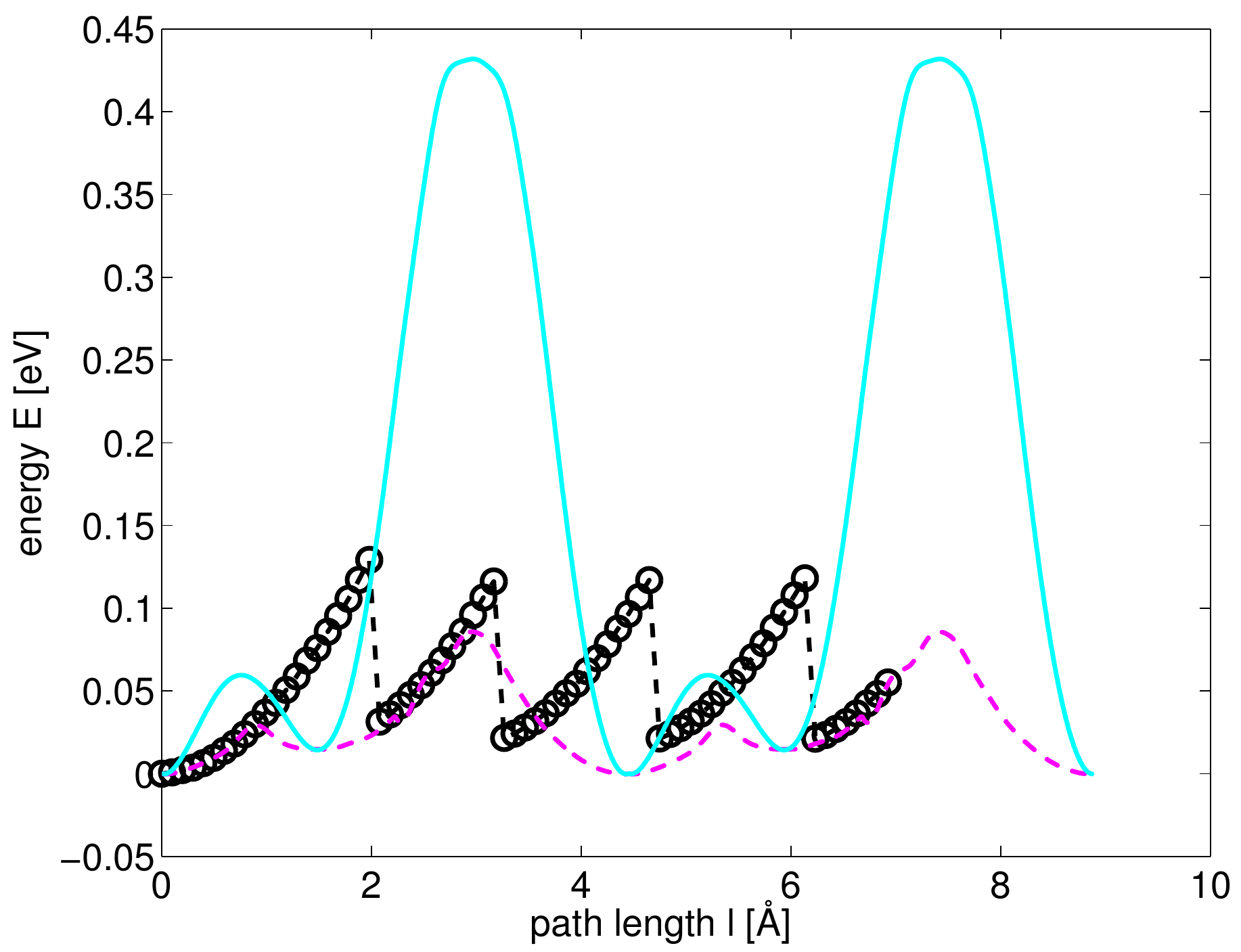}
\caption{A comparison between our friction model and a shearing simulation for the periodic $30^\circ$ sliding path at zero load. Two periods of the quasi static model are shown. The solid (cyan) line and the dashed (magenta) line give the unrelaxed and relaxed energy curves for the quasi-static model, as seen also in figure~\ref{fig:2}. The black circles are fully relaxed data points of the shearing calculation which shows periodic behavior after the first slip with a considerably shorter period than the quasi-static calculation.}
\label{fig:shear_30}
\end{figure}

For the $30^\circ$ (see figure~\ref{fig:shear_30}) sliding direction the situation is different, as the shearing appears to become periodic only after the first slip with a much shorter period than in our quasi static model. Three small slips of roughly equal size happen during a sliding distance of $\sim\unit[4.2]{\AA}$ (from $l\sim\unit[2.0]{\AA}$ to $l\sim\unit[6.2]{\AA}$ in figure~\ref{fig:shear_30}), which is comparable to one period in our quasi-static model of $\sim\unit[4.4]{\AA}$ where one small and one large slip occur. While the predicted slipping process is different the energetics are in excellent agreement, with $E_{\mathrm{qs}}(30^\circ)=\unit[286]{meV}$ being lost per period in our model and $E_{\mathrm{shear}}(30^\circ)=\unit[287]{meV}$ dissipated in three slips during shearing. Using the model of Zhong and Tom\'{a}nek underestimates the energy severely again at $E_{\mathrm{ZT}}(30^\circ)=\unit[104]{meV}$.

This good agreement of our quasi-static results with the shearing calculations adds a convincing argument in favor of our model and justifies the seemingly \textit{ad-hoc} application of unrelaxed energy surfaces. As the periodic paths that were tested in this way turn out to bound the friction force from below and above (see section~\ref{sec:Res}), it is reasonable to assume that the method will also yield plausible results for aperiodic paths for which the shearing calculations become extremely time consuming. The shearing was carried out at constant distance and not constant load, which would allow to adjust the distance between the slabs. This would decrease the pinning and lower the energy barrier, bringing $E_{\mathrm{shear}}$ even closer to the values obtained with our quasi-static model. We want to mention that our energies are calculated at \unit[0]{K}; finite temperature would of course reduce the effective potential corrugation so that a slip process could occur earlier, but this is equally true for both the quasi-static and shear model.

We want to point out, however, that the specific nature of the stick-slip process shown in the shearing calculation is not directly reproduced by our proposed quasi static model (see figures~\ref{fig:3} and~\ref{fig:shear_30}). While the period lengths of the stick-slip cycle are in good agreement for the $0^\circ$ direction, they differ for the $30^\circ$ sliding path. Furthermore, the positions of the slip instabilities in the shearing calculations are of course not reproduced in our model, which allocates the slip points to the maxima of the unrelaxed energy curve. Thus, our new approach is not describing the ``true'' physical dynamics of the slip system, but rather provides a tool to arrive at much better estimates for the dissipated energy than the established model by Zhong and Tom\'{a}nek~\cite{zhong:90,tomanek:91}, at considerable reduced computational effort compared to shearing calculations. 

For the data containing the relaxed and the unrelaxed energy curves as well as the positions of the minima and maxima we propose two distinctively different ways to calculate the mean friction force denoted by $F^{(1)}$ and $F^{(2)}$. To calculate $F^{(1)}$ we determine the Hellmann-Feynman forces acting on the individual atoms in our \textit{ab-initio} calculation \cite{feynman:39}. We perform an arithmetic average of all force components of the upper slab parallel to the sliding direction following the path described above, e.g., as indicated in Fig.~\ref{fig:2} by the dotted line. The contributions of the unrelaxed and the relaxed forces to $F^{(1)}$ are marked by the superscripts $U$ and $R$ while $N_{\mathrm{U}}$ and $N_{\mathrm{R}}$ are the total numbers of sampling points on the unrelaxed and relaxed energy curves yielding
\begin{equation}
F^{(1)}= \frac{1}{N_\mathrm{U}} \sum_{i=1}^{N_\mathrm{U}} F^{\mathrm{U}}_{i} + \frac{1}{N_\mathrm{R}} \sum_{j=1}^{N_\mathrm{R}} F^{\mathrm{R}}_{j} \quad .
\label{eq:F1}
\end{equation}
In contrast, $F^{(2)}$ is defined via the sum over the gains and losses of energy along the sliding path divided by the path length $l$,
\begin{eqnarray}
F^{(2)} & = & \frac{1}{l}\Bigg[ \sum_{i^{\prime}=1}^{N_{\mathrm{max}}} \left(E^{\mathrm{U}}_{i^{\prime}} - E^{\mathrm{R}}_{i^{\prime}} \right) - \sum_{j^{\prime}=2}^{N_{\mathrm{min}}} \left(E^{\mathrm{U}}_{j^{\prime}} - E^{\mathrm{R}}_{j^{\prime}}\right) - \nonumber \\
 & & - \sum_{i^{\prime}=1}^{N_{\mathrm{max}}} \sum_{j^{\prime}=2}^{N_{\mathrm{max}}} \left(E^{\mathrm{R}}_{i^{\prime}}-E^{\mathrm{R}}_{j^{\prime}}\right) \Bigg] \quad .
\label{eq:F2}
\end{eqnarray}
The index $i^{\prime}$ iterates over all maxima in the unrelaxed energy curve and $j^{\prime}$ covers the minima while the energies are denoted by $E^{\mathrm{U}}$ (unrelaxed) and $E^{\mathrm{R}}$ (relaxed). $N_{\mathrm{min}}$ ($N_{\mathrm{max}}$) is the number of minima (maxima) encountered along the chosen sliding path such that $N_{\mathrm{max}}=N_{\mathrm{min}}-1$, as we always start and terminate our path in a minimum of the unrelaxed energy curve for both approaches.

\section{Results and discussion}
\label{sec:Res}

\begin{figure}[htbp]
\centering
\includegraphics[width=1.0\linewidth]{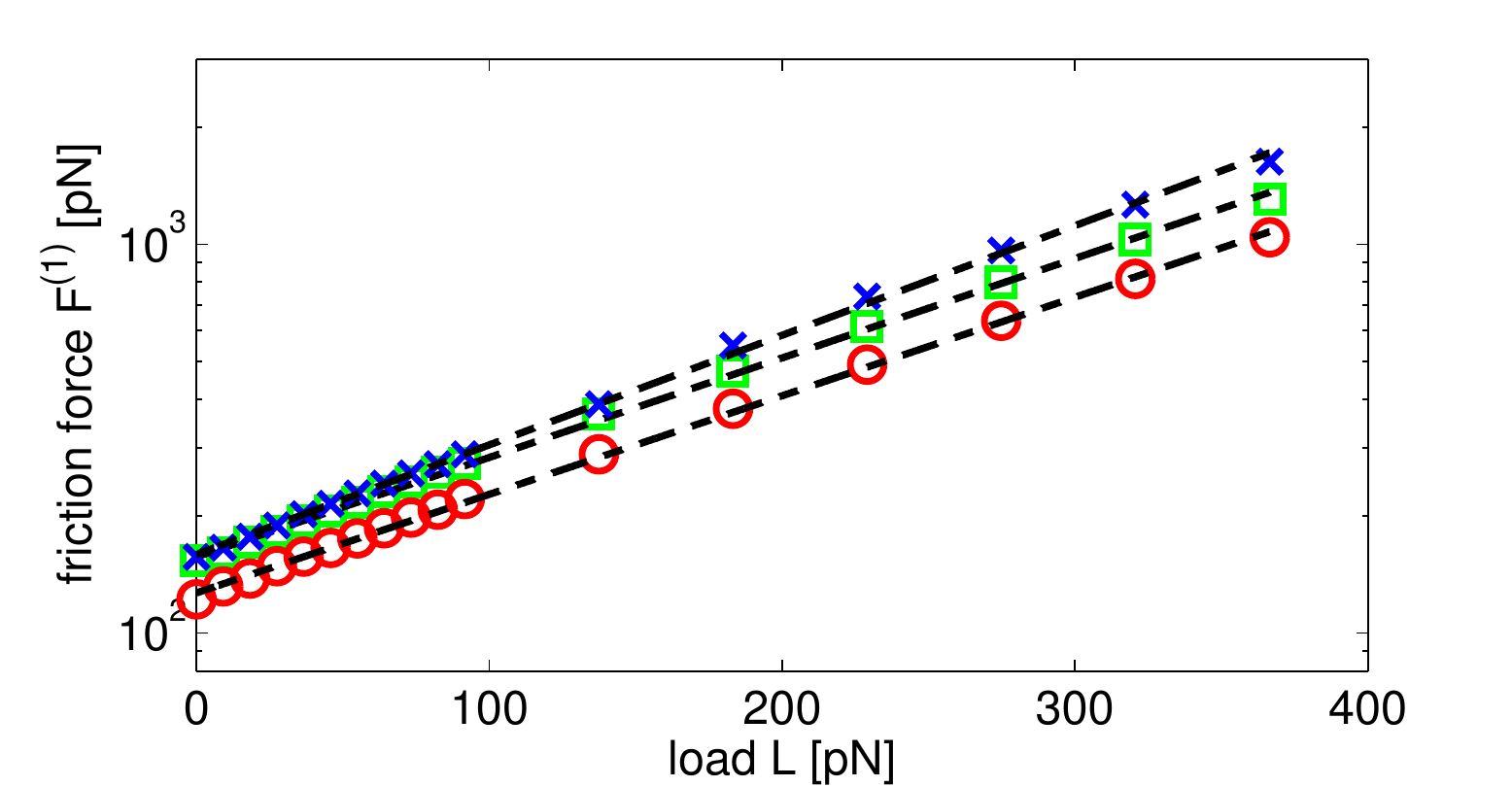}
\caption{Semi-logarithmic plot of the friction force $F^{(1)}$ versus load $L$ for the slip plane of fcc Cu(111). Both periodic paths $0^\circ$ (red circles) and $30^\circ$ (blue crosses) are shown alongside the quasi aperiodic $10^\circ$ path (green squares) which serves as a prototype for all other quasi aperiodic paths. The dashed black lines are linear fits.}
\label{fig:4}
\end{figure}

With the methods described above we calculate the friction-versus-load behavior of fcc Cu(111). All averages are collected on paths of $\unit[1]{\mu m}$ length to get reliable values for the constitutive system parameters and to make sure that in the case of quasi aperiodic paths the whole unit cell becomes sampled. For both definitions of the mean friction force, $F^{(1)}$ and $F^{(2)}$, we observe an exponential friction law along all path directions, see Fig.~\ref{fig:4}, given by
\begin{equation}
\label{eq:exp}
F(L) = F_0 \exp \left( \frac{\mu}{F_0} L \right) \quad .
\end{equation}
This result can be traced back to the exponential form for the binding energy in transition metals e.g.~as given by Pettifor \cite{book:pettifor}. In a very general way the energy of bond breaking follows the \textit{universal binding energy relation} (UBER) which also shows an exponential dependence on distance \cite{rose:83} which in our case depends linearly on load.

The form of the exponent in equation~(\ref{eq:exp}) incorporates the coefficient of friction (COF) $\mu$. Expanding equation~(\ref{eq:exp}) in a Taylor series around $L=0$ we retrieve a kinetic friction law of Derjaguin-form \cite{derjaguin:34b,vernes:12}: linear in load $L$ plus the Derjaguin-offset $F_0$,
\begin{equation}
\label{eq:taylor}
F(L)=F_0 + \mu L + \mathcal{O} \left( L^2 \right) \quad .
\end{equation}
This behavior is well known in lubricated systems where the offset $F_0$ is attributed to adhesion resulting from the lubricant, see e.g. Refs.~\onlinecite{eder:11,vernes:12}. In our dry system the strength of the adhesion, which for frictional purposes acts like an internal load $L_0$, can be estimated by simply lowering one slab down onto the other and registering the forces on the surface atoms. The resulting internal load $L_0$ is the maximum in this force curves depending on the lateral positions of the slabs. It ranges between $\unit[0.8]{nN}$ and $\unit[1]{nN}$ which is of the same order of magnitude as the maximum applied external loads. Of course the approximation of the exponential law with this linear expansion is only valid in the low load regime. For averaged values of all aperiodic paths the relative error reaches $10\%$ at a load of $\unit[90]{pN}$ which corresponds to a pressure of \unit[1.6]{GPa} and should be regarded as the uttermost limit for the Taylor expansion. This low load regime is discussed in more detail later in the paper.

\begin{figure}[Htb]
\centering
\includegraphics[width=1.0\linewidth]{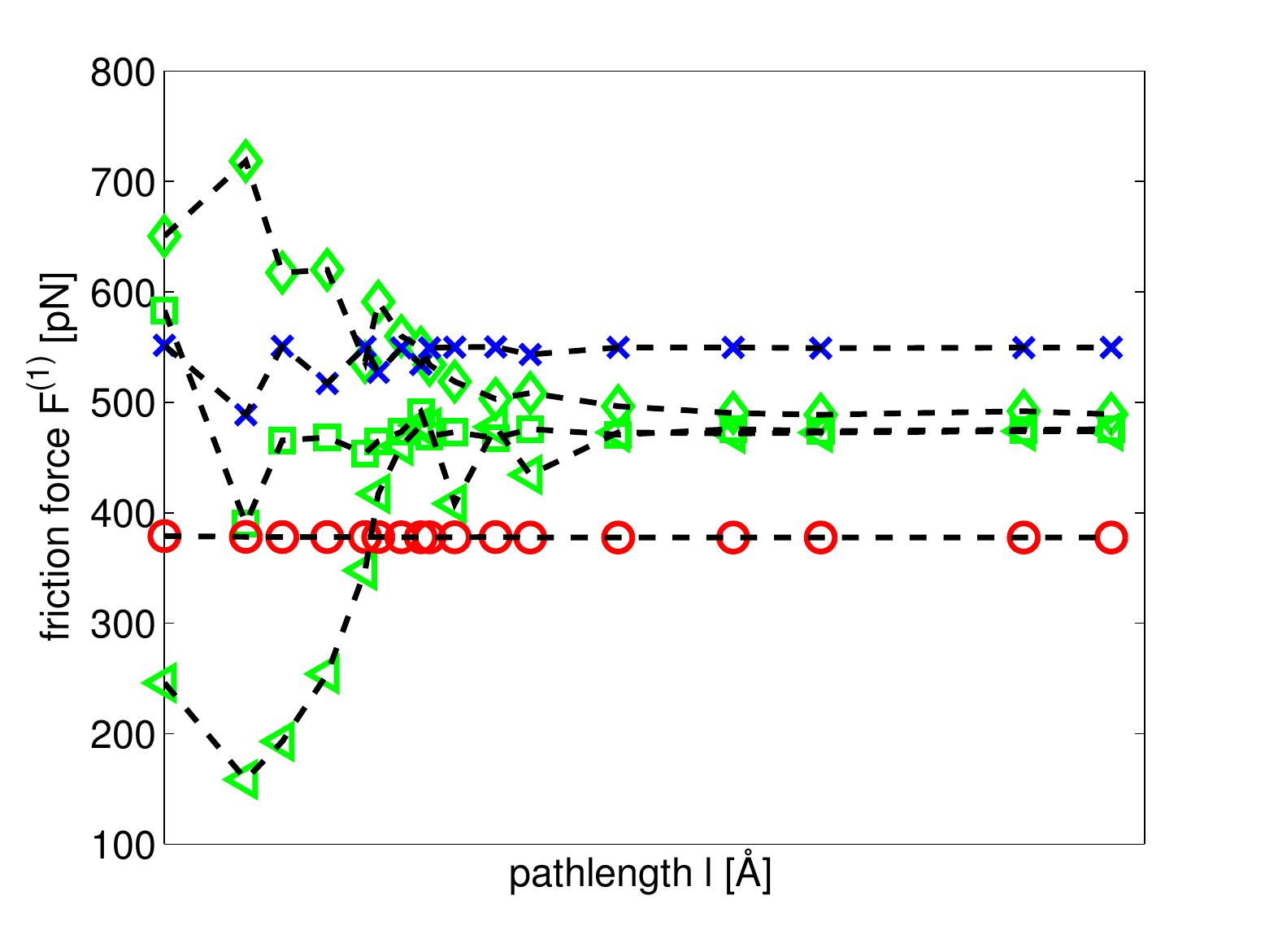}
\caption{Plot of the mean friction force $F^{(1)}$ versus the path length $l$ for a load $L=\unit[183]{pN}$. Both periodic paths $0^\circ$ (red circles) and $30^\circ$ (blue crosses) are shown alongside three quasi aperiodic ones (green diamonds for $-10^\circ$, green triangles for $3^\circ$ and green squares for $10^\circ$). Dashed lines are plotted to guide the eye.}
\label{fig:5}
\end{figure}

While all quasi aperiodic paths investigated for sufficiently long paths converge to the same friction force, there is a clear difference with respect to periodic ones, see Figs.~\ref{fig:4} and \ref{fig:5}. As expected the $30^\circ$ path exhibits the largest friction as for each period it traverses the global maximum of the unrelaxed energy surface created by the on-top position of the contact atoms. This is in contrast to the $0^\circ$ path, which shows the least friction force at zero load and also the slowest increase with rising load.

As the energy loss for the $0^\circ$ path at zero load is underestimated by about 25\% within our model compared to the shearing calculations, the difference between the aperiodic and the $0^\circ$ paths might actually be smaller than predicted. However, even if $F_0(0^\circ)$ is underestimated by our approach, in case that this discrepancy stays approximately the same over the whole range of loads, the predicted coefficients of friction should remain valid anyway. Nevertheless, the existence of a ``hard'' sliding direction at $30^\circ$, as predicted by our model, is supported by the shearing calculations. 

In Fig.~\ref{fig:5} the reason why long paths are required ($\sim~\unit[1]{\mu m}$) becomes apparent. While the friction force for the $0^\circ$ path (red circles) shows essentially no dependence on the length of the path, for the $30^\circ$ one (blue crosses) the friction force oscillates in the beginning before converging to a constant value. This is due to the fact that the summation in equations~(\ref{eq:F1}) and (\ref{eq:F2}) terminates at the last minimum encountered. For $0^\circ$ this happens always after a full period, while along the $30^\circ$ direction one has two minima per period (fcc and the hcp position), so that the total friction force average depends on the termination of the summation. It is obvious that with increasing path length the influence of the termination point becomes less and less important for the calculation of the average. In the quasi aperiodic case, where a short path length would only sample fractions of the energy landscape, it is evident that the friction force only converges for sufficiently long paths.
 
\begin{figure}[Htb]
\centering
\includegraphics[width=1.0\linewidth]{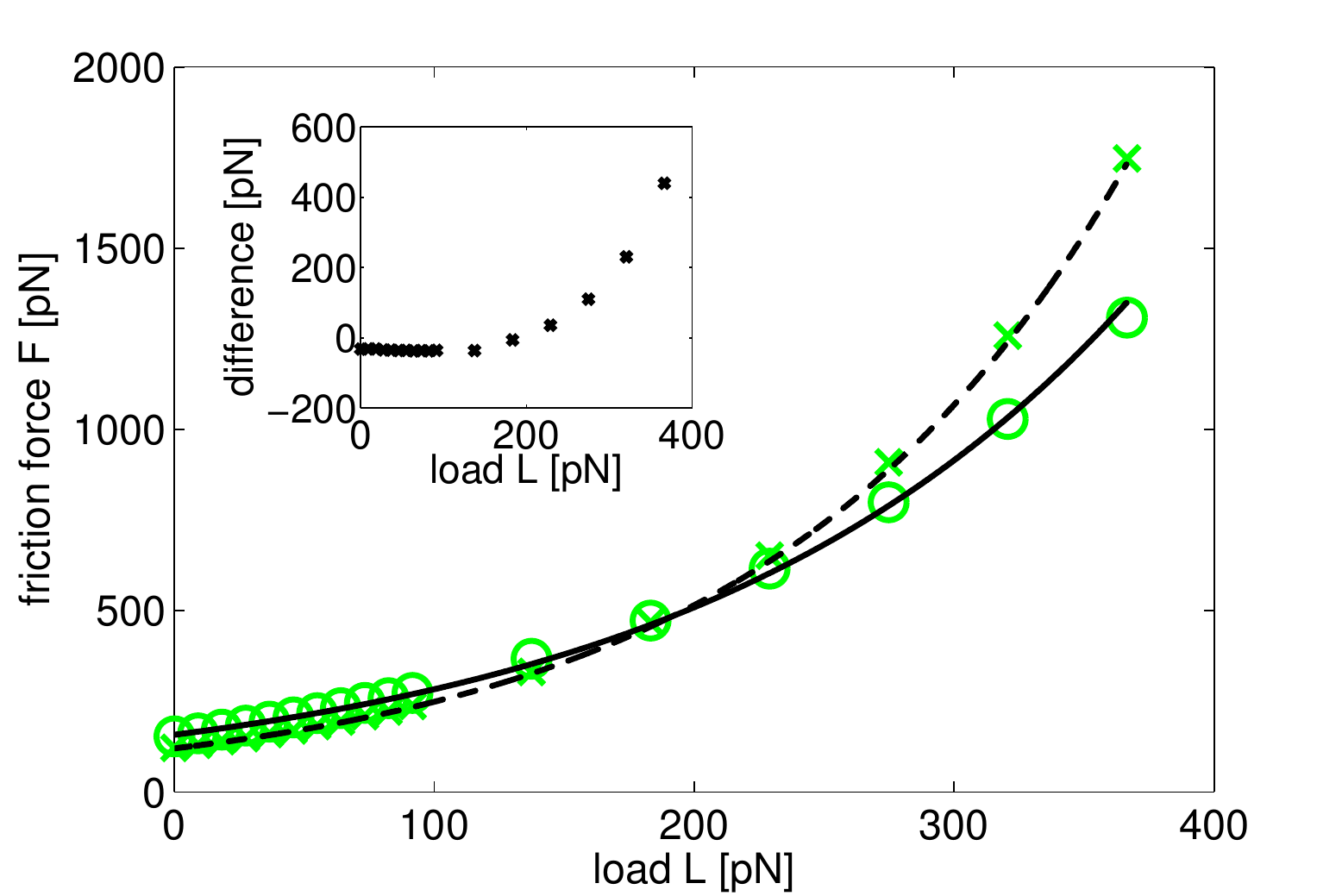}
\caption{Friction forces $F^{(1)}$ (green circles) and $F^{(2)}$ (green crosses) versus load $L$ for fcc Cu(111) slabs sliding along a $10^\circ$ path. The insert (black crosses) shows the difference $F^{(2)}-F^{(1)}$. At low to medium loads the difference is constant and corresponds to the disparity of the friction at zero load $F_0$}
\label{fig:6}
\end{figure}

Both $F^{(1)}$ and $F^{(2)}$ lead to an exponential friction law, however, the calculated values of the Derjaguin-offset $F_0$ and the COF $\mu$, while showing the same trend, differ slightly (see Tab.~\ref{tab:coefficients}). However, this is not surprising given the two different approaches, see equations~(\ref{eq:F1}) and (\ref{eq:F2}). Plotting both friction forces versus load curves for a given sliding path, one finds that $F^{(1)}$ and $F^{(2)}$ do agree well in the low to medium load regime while for loads larger than $L=\unit[150]{pN}$ higher deviations occur, as seen in Fig.~\ref{fig:6}. We attribute this discrepancy to the increasing influence of the reaction forces for large loads which are not considered in the calculation of $F^{(1)}$, but play a role in the dissipation mechanism and hence enter implicitly in $F^{(2)}$.

The largest load considered in our simulations is $\unit[367]{pN}$, given the cross-section area of the unit cell $\unit[5.7]{\AA^2}$, this is equivalent to a static pressure of $\unit[6.4]{GPa}$ which is well beyond any realistic technological applications. Thus, it is important to analyze the low load regime of up to \unit[90]{pN} (\unit[1.6]{GPa}) separately. In this range the exponential law can be approximated reasonably well by a linear relation $F_{\mathrm{lin}}(L)=F_0+\mu_{\mathrm{lin}} L$ so that the COF $\mu_{\mathrm{lin}}$ is given by the slope, see Tab.~\ref{tab:coefficients}. It should be noted that the $\mu_{\mathrm{lin}}$ obtained in this manner are systematically larger than the values of $\mu$, since the slope of a linear fit to a segment of an exponential function is always steeper than the tangent of the function at the beginning of the segment. Since the ultimate tensile strength for annealed copper is approximately $\unit[160]{MPa}$ \cite{yu:04} an uniaxial pressure of more than \unit[1.6]{GPa} will already deform the sample. However, it remains computationally feasible to study extremely high loads because the periodic boundary conditions imposed on our simulations restrict the horizontal movement of the atoms and the fcc (111) symmetry of the sample is preserved at all times.

Depending on direction and method we obtain a COF between $0.6 - 1.46$ which is one order of magnitude larger compared to the MD study by S{\o}rensen et al.~\cite{sorensen:96}. This large difference can be explained at least in part by the small contact size in S{\o}rensen's work that leads to slips mediated by a dislocation mechanism which is suppressed in our model due to the periodic boundary conditions. In 2006 Zhang et al.~\cite{zhang:06} experimentally determined the friction coefficient for dry copper to be $\mu^{\mathrm{nc}}_{\mathrm{exp}}=0.78$ (nanocrystalline sample) and $\mu^{\mathrm{an}}_{\mathrm{exp}}=0.92$ (annealed sample). These values were found in the zero wear regime for a load of $\sim\unit[5]{N}$. The obtained COFs in the low load regime agree fairly well with our values for aperiodic sliding ($\mu^{(1)}=0.94 \pm 0.06$, $\mu^{(2)}=0.88 \pm 0.05$).

\begin{table*}
\caption{\label{tab:coefficients} Derjaguin-offset values $F_0$ and coefficients of friction $\mu$ corresponding to the friction forces  $F^{(1)}$ and $F^{(2)}$, see equations~(\ref{eq:F1}) - (\ref{eq:exp}). The constitutive system parameters for the aperiodic directions are given by the average over all examined quasi aperiodic paths. Experimental data for nanocrystalline (a) and annealed (b) copper are taken from Ref.~\onlinecite{zhang:06}.}
\begin{ruledtabular}
\begin{tabular}{lccccccc}
&$F_0^{(1)}[\mathrm{pN}]$	& $F_0^{(2)}[\mathrm{pN}]$	& $\mu^{(1)}$ & $\mu^{(2)}$ & $\mu^{(1)}_{\mathrm{lin}}$ & $\mu^{(2)}_{\mathrm{lin}}$ & $\mu_{\mathrm{exp}}$\\
$0^\circ$	& 127	& 101	& 0.74 & 0.60 & 1.06 & 0.87 & -	\\
$30^\circ$	& 159	& 109	& 1.03 & 0.91 & 1.45 & 1.46 & -	\\
aperiodic	& $157 \pm 4.4$	& $121 \pm 2.7$	& $0.94 \pm 0.06$ & $0.88 \pm 0.05$	& $1.33 \pm 0.14$ & $1.25 \pm 0.04$ & $0.78^{a}$, $0.92^{b}$ \\
\end{tabular}
\end{ruledtabular}
\end{table*}

\section{\label{sec:Con}Conclusion}
We present a novel approach to construct nanofriction vs.~load curves from total energy landscapes generated by DFT calculations. The method is parameter-free as no external input is needed to compute the friction force since internal relaxations of the system are assumed to dissipate the energy. We are able to study very long sliding paths and find that the friction force on the nanoscale converge for all aperiodic paths to a value between the limits set by two paths along high symmetry directions (Fig.~\ref{fig:5}). Comparisons with more realistic but computationally more demanding shearing calculations, that were carried out for the limiting cases of low and high friction, show that, while the exact dynamics of the sliding system are not reproduced well, the estimation of the energy loss is significantly improved compared to established methods. We define two distinct ways to estimate the mean friction force which both lead to comparable results and yield an exponential friction law which in the low load regime can be expanded to a linear relation of Derjaguin form. These two methods serve primarily as an internal test of consistency of our model, however, since total energies can be calculated with a higher accuracy than forces, we would give preference to $F^{(2)}$ (as given by equation~\ref{eq:F2}). The exponential increase of friction with applied load is comparable to the experimental findings of Gosvami et al.~\cite{gosvami:10} where a strong increase of friction at high loads was found for Au(111) and Cu(100). A similar behavior is found in Fig.~12 of Ref.~\onlinecite{dong:13} by performing classical MD simulations. However, both groups attribute this strong increase in friction to the onset of wear at high loads, which is not considered in our work. We show that we can obtain good linear dependence of the friction force on the load in the low load regime (which is still under a high internal load due to adhesion) reminiscent of the Amontons-Coulomb law and also often found on the nanoscale \cite{he:99,mueser:01,gao:04}. The calculated coefficients of friction fit well to the measured values in macroscopic experiments, but there remains the possibility of a coincidental agreement which would require further studies, both theoretically and experimentally, to be ruled out.

\section*{Acknowledgements}
This work was funded by the ``Austrian COMET-Program'' (project
XTribology, no. 824187) via the Austrian Research Promotion Agency (FFG) and the Province of Nieder\"osterreich, Vorarlberg and Wien and has been carried out within the ``Excellence Centre of Tribology'' (AC2T research GmbH) and at Vienna University of Technology. P.M., J.R., and G.F. acknowledge the support by the Austrian science fund (FWF) [SFB ViCoM F4109-N13]. The authors want to thank Florian Mittendorfer for fruitful discussions and also appreciate the ample support of computer resources by the Vienna Scientific Cluster (VSC). Fig.~\ref{fig:0} in this paper was created with the help of the VESTA code~\cite{vesta:11}.

\bibliography{../../PhD_Papers/Bib}

\end{document}